\documentclass[conference]{IEEEtran}
\IEEEoverridecommandlockouts
\usepackage{cite}
\usepackage{amsmath,amssymb,amsfonts}
\usepackage{algorithm}
\usepackage{algpseudocode}
\usepackage{graphicx}
\usepackage{tikz}
\usepackage{textcomp}
\usepackage{xcolor}
\usepackage{booktabs}
\usepackage{multirow}
\usepackage{url}
\usepackage[hidelinks]{hyperref}
\hypersetup{pdftitle={Where Should the Vectors Live? Locality-Aware Retrieval for RAG on Geo-Distributed SQL}, pdfauthor={Sharath Chandra Kampili, Abhilash Rao Mesala, Ramakrishna Kommineni}}
\newcommand{\NumPassages}{20{,}963}
\newcommand{\NumEast}{7{,}710}
\newcommand{\NumWest}{6{,}599}
\newcommand{\NumEU}{6{,}654}
\newcommand{\NumBenchQ}{600}
\newcommand{\NumBenchRuns}{10{,}800}
\newcommand{\NumQSample}{3{,}000}
\newcommand{\ExactHitTen}{73.7\%}
\newcommand{\ExactHitTenBench}{72.0\%}
\newcommand{\CalEastPoint}{0.4}
\newcommand{\CalWestPoint}{67.2}
\newcommand{\CalEUPoint}{91.3}

\newcommand{\LatCenEU}{372.6}

\newcommand{\HitCen}{65.3\%}
\newcommand{\RecCen}{90.4\%}

\newcommand{\LatFrEU}{6.4}

\newcommand{\LatGloEU}{6.7}

\newcommand{\HitGlo}{66.3\%}
\newcommand{\RecGlo}{90.5\%}

\newcommand{\LatFanEast}{500.9}

\newcommand{\LatFanEU}{716.6}

\newcommand{\LatParEast}{285.4}

\newcommand{\LatParEU}{429.0}

\newcommand{\HitPar}{70.5\%}
\newcommand{\RecPar}{97.3\%}

\newcommand{\HitLoc}{25.8\%}

\newcommand{\HitHomeLoc}{77.3\%}

\newcommand{\CentralEUoverRTT}{4.2}
\newcommand{\CentralWestoverRTT}{4.2}

\newcommand{\LocalMin}{6.4}
\newcommand{\LocalMax}{7.4}
\newcommand{\ParOverLocMin}{38}
\newcommand{\ParOverLocMax}{67}
\newcommand{\MaxSpeedup}{112$\times$}
\newcommand{\FrLagSec}{4.24}
\newcommand{\BeamDefRec}{89.9\%}
\newcommand{\BeamDefLat}{6.2}
\newcommand{\BeamMaxRec}{99.2\%}
\newcommand{\BeamMaxLat}{24.5}

\newcommand{\BeamMaxHit}{73.7\%}
\newcommand{\BeamDefHit}{67.3\%}

\newcommand{\IngCenWest}{336.7}
\newcommand{\IngCenEU}{457.5}
\newcommand{\IngPartEast}{4.2}

\newcommand{\IngPartEU}{2.8}
\newcommand{\IngRbrEast}{96.5}
\newcommand{\IngRbrWest}{143.7}

\newcommand{\IngGloEast}{808.2}

\newcommand{\IngGloEU}{1010.7}

\newcommand{\SgLatParNine}{379.3}
\newcommand{\SgHitParNine}{70.5\%}

\newcommand{\SgHitLocNine}{69.6\%}

\newcommand{\LarrTau}{0.611}
\newcommand{\LarrTauPct}{70}

\newcommand{\LarrHitOne}{75.5\%}

\newcommand{\LarrLatNine}{64.1}

\newcommand{\LarrHitNine}{70.7\%}
\newcommand{\LarrEscNine}{14.6\%}

\newcommand{\LarrHitUni}{43.4\%}
\newcommand{\LarrEscUni}{30.0\%}
\newcommand{\LarrSpeedupNine}{5.9$\times$}

\newcommand{\LiveNQ}{300}

\newcommand{\LiveLatNine}{59.0}
\newcommand{\LiveHitNine}{70.3\%}
\newcommand{\LiveEscNine}{12.5\%}

\newcommand{\AucDone}{0.78}

\newcommand{\OrLatNine}{46.2}
\newcommand{\OrHitNine}{76.7\%}

\newcommand{\OrHitUni}{72.8\%}
\newcommand{\CIHitDiff}{[-1.4, +1.9]}
\newcommand{\CISpeedup}{5.1--6.9}

\newcommand{\AucSkA}{0.83}
\newcommand{\AucSkB}{0.88}
\newcommand{\AucSkC}{0.88}
\newcommand{\AucDOneTest}{0.79}
\newcommand{\AucSketchOnly}{0.86}
\newcommand{\SkGainMax}{3.3}

\newcommand{\AucLocalFeat}{0.80}
\newcommand{\RepLocAbs}{0.8}
\newcommand{\RepRemRel}{0.5\%}
\newcommand{\RepN}{200}
\newcommand{\NineCenEU}{379.6}
\newcommand{\NineCenWest}{283.1}

\newcommand{\NineAmpEU}{4.1}
\newcommand{\NineAmpWest}{4.1}
\newcommand{\OneAmpEUsub}{4.2}
\newcommand{\NineLocMin}{11.1}
\newcommand{\NineLocMax}{18.4}
\newcommand{\NineParMin}{291.9}
\newcommand{\NineParMax}{435.8}
\newcommand{\NineLarrHit}{73.7\%}
\newcommand{\NineParHit}{75.0\%}
\newcommand{\NineLarrLat}{75.8}
\newcommand{\NineParLat}{391.2}
\newcommand{\NineSpeedup}{5.2$\times$}
\newcommand{\OneLarrHitSub}{71.8\%}
\newcommand{\OneParHitSub}{73.0\%}
\newcommand{\OneSpeedupSub}{5.9$\times$}
\newcommand{\NineIngLocMin}{9.4}
\newcommand{\NineIngLocMax}{11.7}
\newcommand{\NineIngCenEU}{465.0}
\newcommand{\CcLocPeak}{299}
\newcommand{\CcSOneSixteen}{42.4}

\newcommand{\CcLarrSixteen}{123}
\newcommand{\CcLarrSixteenPfifty}{32.3}
\newcommand{\CcLarrSixteenPnf}{714}

\newcommand{\CcSOneOne}{2.8}
\newcommand{\CcLarrUniSixteen}{81}
\newcommand{\CcSOneClientsHundred}{36}
\newcommand{\MlExact}{86.9\%}
\newcommand{\MlNQ}{600}
\newcommand{\MlCenEU}{373.8}
\newcommand{\MlAmp}{4.2}
\newcommand{\MlLocMin}{6.5}
\newcommand{\MlLocMax}{7.0}
\newcommand{\MlHitCen}{83.7\%}
\newcommand{\MlHitPar}{85.3\%}
\newcommand{\MlHitHomeLoc}{88.3\%}
\newcommand{\MlLarrHit}{83.9\%}
\newcommand{\MlParHit}{84.0\%}
\newcommand{\MlLarrLat}{110.7}
\newcommand{\MlParLat}{388.3}
\newcommand{\MlSpeedup}{3.5$\times$}
\newcommand{\MlLarrEsc}{27.3\%}
\newcommand{\MlAucDone}{0.88}
\newcommand{\MlAucSketch}{0.92}
\newcommand{\MlCalBest}{86.65\%}
\newcommand{\MlCalFan}{86.67\%}
\newcommand{\MlTauPct}{40}
\newcommand{\MlMedHit}{83.6\%}
\newcommand{\MlMedLat}{78.9}
\newcommand{\MlMedSpeedup}{4.9$\times$}
\newcommand{\LarrSaveMs}{315}
\newcommand{\ScMultA}{3.2}
\newcommand{\ScMeasA}{3.0}
\newcommand{\ScEUA}{284.0}
\newcommand{\ScLocA}{8.2}
\newcommand{\ScMultB}{4.2}
\newcommand{\ScMeasB}{4.0}
\newcommand{\ScEUB}{375.0}

\newcommand{\ScMultC}{4.2}
\newcommand{\ScMeasC}{4.0}
\newcommand{\ScEUC}{376.4}
\newcommand{\ScLocC}{10.6}
\newcommand{\ScChunks}{102{,}706}
\newcommand{\KmeansSec}{5.5}
\newcommand{\CwNQ}{600}
\newcommand{\CwType}{e2-standard-2}
\newcommand{\CwRttWest}{66}
\newcommand{\CwRttEU}{100}

\newcommand{\CwCenEU}{424}

\newcommand{\CwAmpWest}{4.3}
\newcommand{\CwAmpEU}{4.2}
\newcommand{\CwLocMin}{7.1}
\newcommand{\CwLocMax}{8.7}
\newcommand{\CwLarrHit}{70.4\%}
\newcommand{\CwFanHit}{70.2\%}
\newcommand{\CwLarrLat}{74.8}
\newcommand{\CwFanLat}{457.6}
\newcommand{\CwSpeedup}{6.1$\times$}

\newcommand{\CwLiveN}{300}
\newcommand{\CwLiveHit}{70.2\%}
\newcommand{\CwLiveLat}{70.0}
\newcommand{\CwFanPnfWest}{912}
\newcommand{\CwFanPfWest}{487}

\newcommand{\WnNQ}{200}
\newcommand{\WnAmpMin}{4.0}
\newcommand{\WnAmpMax}{4.2}

\newcommand{\WnSpeedJit}{6.0$\times$}

\newcommand{\WnSpeedNear}{4.2$\times$}

\newcommand{\WnJitPct}{10}
\newcommand{\WnHitSame}{identical}

\begin{document}

\title{Where Should the Vectors Live? Locality-Aware Retrieval for RAG on Geo-Distributed SQL}

\author{\IEEEauthorblockN{Sharath Chandra Kampili}
\IEEEauthorblockA{\textit{Independent Researcher} \\
United States \\
sharathchandra.kampili@ieee.org}
\and
\IEEEauthorblockN{Abhilash Rao Mesala}
\IEEEauthorblockA{\textit{Independent Researcher} \\
United States \\
mesala.arr@gmail.com}
\and
\IEEEauthorblockN{Ramakrishna Kommineni}
\IEEEauthorblockA{\textit{Independent Researcher} \\
United States \\
ramakrishna.kommineni4@gmail.com}
}

\maketitle

{\let\thefootnote\relax\footnotetext{\scriptsize Accepted for publication in the 2026 International Conference on Artificial Intelligence Centric Systems, Services, and Applications (AISSA). Author accepted version. \copyright~2026 IEEE. Personal use of this material is permitted. Permission from IEEE must be obtained for all other uses, in any current or future media, including reprinting/republishing this material for advertising or promotional purposes, creating new collective works, for resale or redistribution to servers or lists, or reuse of any copyrighted component of this work in other works.}}

\begin{abstract}
Enterprise retrieval-augmented generation (RAG) increasingly runs vector search inside geo-distributed SQL databases, where the corpus is fresh, transactional, and subject to data-residency rules. Where the vectors live relative to their users then decides retrieval latency, freshness, and compliance. We compare five placement strategies on a three-region deployment of a commercially available geo-distributed SQL database with emulated wide-area latency and its native vector index: a centralized table, follower reads, a global table, and home-region partitioning searched globally or locally. The workload is \NumPassages{} SQuAD passages with real questions. Approximate nearest neighbor search amplifies network distance: a remote top-10 query costs \CentralEUoverRTT{} round trips beyond its local cost, so medians span \LocalMin\,ms to over 700\,ms (\MaxSpeedup) across strategies that differ only in placement. Local strategies pay elsewhere, in staleness, write latency, or missed answers. We propose LARR, a locality-aware retrieval router that searches the home partition first and escalates to a cross-region fan-out only when the best local distance exceeds a calibrated threshold. With 90\% of questions issued from the answer's home region, LARR matches the fan-out's hit@10 (\LarrHitNine{} vs.\ \SgHitParNine; 95\% CI of the difference \CIHitDiff{} points) at \LarrSpeedupNine{} lower mean latency. The result holds live on held-out questions, with production-style three-node-per-region replication, and on three real cloud regions (\CwSpeedup). With a second, transformer-based embedding model, LARR matches the fan-out at \MlSpeedup{} lower latency, and its throughput advantage persists under concurrent load. A replicated centroid sketch (LARR-S) sharpens escalation when locality is weaker. We also report a pitfall with partitioned vector indexes.

\end{abstract}


\section{Introduction}
Retrieval-augmented generation (RAG) grounds a large language model (LLM) in documents retrieved at query time~\cite{lewis2020rag,gao2023ragsurvey}. In enterprise deployments the retrieval corpus is rarely a static file: it is operational data---tickets, contracts, product records, customer conversations---that is written continuously, protected by access rules, and increasingly subject to data-residency obligations such as the EU General Data Protection Regulation (GDPR)~\cite{gdpr2016}. These pressures are pushing vector search into the operational database itself. Distributed SQL systems now ship native, transactionally consistent vector indexes. Some store a SPANN-style~\cite{chen2021spann,xu2023spfresh} hierarchical partition tree as ordinary replicated key ranges and compress vectors with quantization such as RaBitQ~\cite{gao2024rabitq}.

Once vectors live in a geo-distributed database, a question that vector-database benchmarks do not address becomes central: \emph{where should the vectors live relative to the users who query them?} Multi-region SQL exposes several table localities---a table homed in one region, a table replicated for low-latency reads everywhere, a table partitioned row-by-row to each row's home region---and each composes with an approximate nearest neighbor (ANN) index differently. A point lookup pays a wide-area round trip (RTT) at most once; an ANN query walks an index tree and then re-ranks candidates, issuing several dependent reads. If those reads cross an ocean over the wide-area network (WAN), retrieval can consume a large share of an interactive RAG latency budget before the LLM generates a single token.

This paper makes the following contributions:
\begin{enumerate}
\item A \textbf{taxonomy of five vector-placement strategies} on multi-region distributed SQL (Section~\ref{sec:placement}), with the consistency, freshness, residency, and write-cost trade-off each implies.
\item A \textbf{reproducible three-region testbed} built on a commercially available geo-distributed SQL database with emulated inter-region RTTs and a public question-answering workload (SQuAD~\cite{rajpurkar2016squad}: \NumPassages{} passages, real questions with known answer passages).
\item An \textbf{empirical characterization} (Section~\ref{sec:results}) of retrieval latency per client region, retrieval quality, ingest latency, and freshness for each strategy, including throughput under concurrent load. A remote ANN query costs \CentralEUoverRTT{} inter-region RTTs beyond its local cost, so placements that look equivalent for point reads differ by up to \MaxSpeedup{} for vector retrieval.
\item \textbf{LARR}, a \emph{Locality-Aware Retrieval Router} (Section~\ref{sec:larr}). LARR searches the client's home partition first and escalates to a cross-region fan-out only when the local result is not confident. When 90\% of questions are local, it matches the quality of a global fan-out at \LarrSpeedupNine{} lower mean latency, and a live run on held-out questions reproduces these figures. We also quantify where its distance-based signal breaks down, and we show that a small replicated centroid sketch (LARR-S) sharpens it on held-out questions.
\item \textbf{Robustness checks}: bootstrap confidence intervals, a repeat run, a production-style three-nodes-per-region cluster (Section~\ref{sec:ninenode}), and a second, transformer-based embedding model (Section~\ref{sec:minilm}).
\item A practical \textbf{pitfall} with partitioned vector indexes, and a self-retrieval check that detects it (Section~\ref{sec:discussion}).
\end{enumerate}
The code, testbed scripts, and raw measurements are available from the author on request.

\section{Background and Related Work}
\label{sec:related}
\textbf{ANN indexes.} Graph indexes such as HNSW~\cite{malkov2018hnsw} and DiskANN~\cite{subramanya2019diskann}, and partition (inverted-file) indexes such as FAISS IVF~\cite{johnson2019faiss} and SPANN~\cite{chen2021spann}, trade recall for latency through a search-width parameter. SPFresh~\cite{xu2023spfresh} supports in-place updates to a SPANN-style index. Transactional databases have adapted this design to a key-value store: partitions and their centroids are stored as rows, partitions split and merge in the background, and optional \emph{prefix columns} give each prefix value (e.g., a tenant or a region) its own partition tree.

\textbf{Vector data management.} Purpose-built systems such as Milvus~\cite{wang2021milvus} and hybrid engines such as AnalyticDB-V~\cite{wei2020analyticdbv} focus on throughput and recall within a single site. Filtered ANN methods such as Filtered-DiskANN~\cite{gollapudi2023filtered} and ACORN~\cite{patel2024acorn} restrict results to rows that match a predicate, and RAG serving systems such as RAGCache~\cite{jin2024ragcache} cut latency by caching within one site. These evaluations rarely model wide-area placement, which is the focus here. Our local-only strategy is, in effect, a filter on the region attribute, but its cost is governed by network distance rather than filter selectivity.

\textbf{Geo-distributed databases.} Spanner~\cite{corbett2012spanner} and the systems that followed it replicate data with consensus and let operators control where leaseholders and replicas live. Their multi-region abstractions---tables pinned to a home region, row-level regional partitioning, \emph{global} tables readable locally everywhere, and bounded-staleness \emph{follower reads}---were designed and evaluated mainly for online transaction processing (OLTP) point operations.

\textbf{Retrieval routing.} Distributed information retrieval has long studied \emph{resource selection}, choosing which collections to query~\cite{callan2000dir}. For RAG, RAGRoute learns which federated sources to query~\cite{dhasade2025ragroute}, Self-RAG and Adaptive-RAG decide \emph{whether} to retrieve~\cite{asai2024selfrag,jeong2024adaptiverag}, and EdgeRAG reduces on-device retrieval latency through online indexing~\cite{seemakhupt2024edgerag}. LARR differs in signal and scope: it routes between the local and global scope of a \emph{single} consistent database, using only the ANN distance the database already returns, with no learned model.

\section{Vector Placement Strategies}
\label{sec:placement}
We consider a corpus of passage embeddings in a database spanning regions $\mathcal{R}$; each passage has a \emph{home region} (where it was created or must reside), and each query arrives at a gateway in region $g$. Let $\mathrm{RTT}(a,b)$ denote the round-trip time between regions. Table~\ref{tab:strategies} summarizes five strategies.

\textbf{S1 Centralized} (table pinned to one region). Every passage and the ANN index have their leaseholder in one region. Reads are strongly consistent; a remote query pays $\mathrm{RTT}(g,\text{home})$ for \emph{each} dependent read the index traversal issues.

\textbf{S2 Follower reads.} The same table, queried at a bounded-staleness timestamp a few seconds in the past. Any replica in $g$ can serve the query locally, but new passages become visible only after the closed timestamp advances.

\textbf{S3 Global table.} Reads are consistent \emph{and} local in every region; writes pay a commit wait so that all replicas can serve present-time reads. Every region holds a full copy, which conflicts with residency requirements.

\textbf{S4 Home-partitioned, global fan-out.} Each passage is stored only in its home region, either as a row-level regionally partitioned table whose vector index is partitioned by a region column, or as one region-pinned table per region. Both give the same leaseholder and replica placement. A complete top-$k$ answer must search every region's partition and merge the results, either in one SQL statement (S4a) or as concurrent per-region queries merged by the client (S4b).

\textbf{S5 Home-partitioned, local only.} The query searches only the gateway region's partition. It is fully local but blind to passages homed elsewhere.

\begin{table*}[t]
\centering
\caption{Placement strategies and their properties.}
\label{tab:strategies}
\footnotesize
\setlength{\tabcolsep}{8pt}
\begin{tabular}{@{}lcccc@{}}
\toprule
Strategy & Remote-read cost & Freshness & Residency & Write cost \\
\midrule
S1 Centralized & WAN per index read & fresh & one region & local at home \\
S2 Follower reads & local & stale (s) & copies everywhere & local at home \\
S3 Global & local & fresh & copies everywhere & commit wait \\
S4 Partitioned fan-out & WAN to every region & fresh & home only & local$^\dagger$ \\
S5 Partitioned local & local & fresh & home only & local$^\dagger$ \\
\bottomrule
\end{tabular}
\\[2pt]{\footnotesize $^\dagger$With row-level regional partitioning, a cross-region uniqueness check is added unless the key includes the region.}
\end{table*}

\section{LARR: Locality-Aware Retrieval Router}
\label{sec:larr}
S5 is fast but incomplete; S4 is complete but slow. Locality in real workloads is high but not perfect: most users ask about data homed near them, while some questions need data from elsewhere. LARR exploits this skew. It first runs the top-$k$ ANN query on the gateway's home partition. If the best cosine distance $d_1$ is at most a threshold $\tau$, the local answer is returned. Otherwise LARR escalates to the global fan-out query (S4), whose results contain any better remote passages.

With escalation probability $p_e(\tau)$, the expected latency of the sequential variant is $\mathbb{E}[T] = T_{\text{loc}} + p_e(\tau)\,T_{\text{glob}}$. A \emph{hedged} variant issues both queries at once and returns the local result if it is confident on arrival, otherwise the global result; this caps escalated latency at $\max(T_{\text{loc}},T_{\text{glob}})$ in exchange for always paying the cross-region work~\cite{dean2013tail}. The threshold $\tau$ is set from the empirical distribution of local top-1 distances: choosing the $p$-th percentile targets an escalation rate of about $1-p$. LARR needs no model, no extra metadata, and no change to the schema.

\textbf{Choosing and maintaining $\tau$.} Because $\tau$ is a percentile of local top-1 distances, it can be set from unlabeled traffic: an operator who can afford an escalation rate $e$ sets $\tau$ to the $(1-e)$-quantile of recent $d_1$ values. When a small labeled or judged sample is available, $\tau$ can instead be set to the lowest-latency value that matches the fan-out's quality, as we do in Section~\ref{sec:results}. As locality shifts (new tenants, migrations, time-of-day effects), the $d_1$ distribution shifts with it. Re-estimating the quantile over a sliding window holds the escalation rate, and \emph{shadow-sampling} a small random fraction of queries through the fan-out measures, without labels, how often escalation would have changed the top result. That rate is an online proxy for the quality gap.

\textbf{Residency and escalation.} LARR stores every passage only in its home region, so it satisfies residency rules on \emph{storage}. An escalation still \emph{reads} remote partitions, which may conflict with rules on cross-border \emph{access}. Three deployment modes cover the common cases: (i) \emph{policy-scoped escalation}, where the fan-out is restricted to the regions the requesting principal may read; (ii) \emph{identifier-only escalation}, where remote partitions return only passage identifiers and distances, and text is fetched only where policy allows; and (iii) \emph{local-only with disclosure}, where queries whose $d_1$ exceeds $\tau$ are answered locally and flagged to the application as possibly incomplete. Because every cross-region read is a discrete escalation event, LARR also makes cross-border access auditable, which an always-global fan-out does not.

\textbf{LARR-S: sketch-assisted escalation.} The local top-1 distance only says how good the local answer is. It does not say whether a better answer exists elsewhere. LARR-S adds that evidence cheaply. Each region publishes a \emph{sketch} of $K$ k-means centroids of its own embeddings (256 centroids $\times$ 256 dimensions $\approx$ 256\,KB per region at 32-bit precision), which is replicated to every region. For a query $q$ at gateway $g$, the gateway computes the sketch margin
\[ m(q,g) = \max_{c \in C_g} \langle q, c\rangle - \max_{r \neq g}\, \max_{c \in C_r} \langle q, c\rangle \]
in memory, with no network round trip. A logistic model over $(d_1, m)$ then estimates the probability that the answer is homed locally, and LARR-S escalates when that probability falls below a threshold. Only centroids, each the mean of tens of passages, leave a region. Deployments with strict residency rules can enforce a minimum cluster size or keep the sketch off. Compared with LARR, LARR-S costs three things: a periodic k-means refresh per region (\KmeansSec\,s for 256 centroids over a region's \NumEast{} vectors on our host), replication of the sketch, and a small labeled sample to fit the two-feature logistic model. LARR needs none of these. We therefore recommend starting with LARR and adding LARR-S only when locality is moderate ($\sigma \le 0.8$), which is where it helps (Section~\ref{sec:results}).

\section{Experimental Setup}
\label{sec:setup}
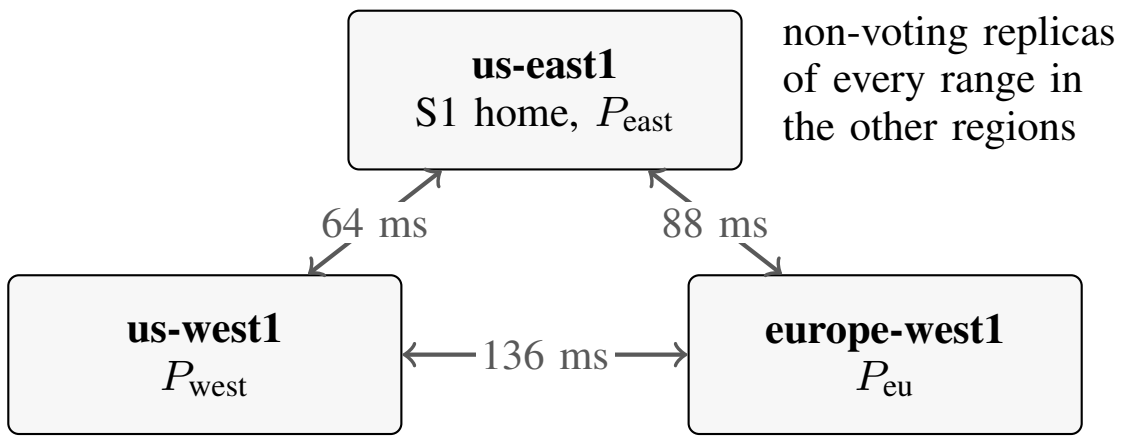
\begin{figure}[t]
\centering
\begin{tikzpicture}[
  font=\footnotesize,
  region/.style={draw, rounded corners=2pt, minimum width=2.6cm, minimum height=1.05cm, align=center, fill=gray!6},
  link/.style={<->, thick, gray!70!black},
  lbl/.style={fill=white, inner sep=1pt, font=\footnotesize}]
\node[region] (E) at (0,0) {\textbf{us-east1}\\ S1 home, $P_{\text{east}}$};
\node[region] (W) at (-2.25,-1.75) {\textbf{us-west1}\\ $P_{\text{west}}$};
\node[region] (U) at (2.25,-1.75) {\textbf{europe-west1}\\ $P_{\text{eu}}$};
\draw[link] (E) -- node[lbl]{64 ms} (W);
\draw[link] (E) -- node[lbl]{88 ms} (U);
\draw[link] (W) -- node[lbl]{136 ms} (U);
\node[align=left, font=\footnotesize, anchor=west] at (1.45,0.05) {non-voting replicas\\of every range in\\the other regions};
\end{tikzpicture}
\caption{Testbed: one node (and gateway) per region; inter-region traffic passes through user-space delay proxies emulating the RTTs shown.}
\label{fig:testbed}
\end{figure}

\textbf{Testbed.} We run a commercially available, Spanner-style geo-distributed SQL database (recent 2026 release; we do not name the product) on one host (2 vCPUs, 8\,GB RAM). Three nodes (Fig.~\ref{fig:testbed}) carry localities \texttt{us-east1}, \texttt{us-west1}, and \texttt{europe-west1}. The sandbox kernel lacks \texttt{netem}, so we emulate the WAN in user space. Using per-locality advertised addresses, each node gives a direct address to peers in its own region and a delay-proxy address to peers in each other region. The proxy adds half the region-pair RTT in each direction. The emulated RTTs are 64\,ms (us-east1--us-west1), 88\,ms (us-east1--europe-west1), and 136\,ms (us-west1--europe-west1), chosen to approximate published public-cloud inter-region latencies. Calibration point reads against the us-east1-homed table take \CalWestPoint\,ms from us-west1 and \CalEUPoint\,ms from europe-west1, versus \CalEastPoint\,ms locally. With one node per region, the default zone-level survivability setting (three voters in the home region) cannot be placed. We therefore configure one voting replica (and leaseholder) in the home region and non-voting replicas in the two other regions. This preserves the read-path behavior of every locality and omits the roughly 1\,ms of intra-region replication a production deployment would add to writes. The emulator adds a fixed delay in each direction. It does not model packet loss or bandwidth limits, so Section~\ref{sec:wan} repeats the core measurements on three real cloud regions and under jitter and shorter RTTs.

\textbf{Configuration.} Each node runs with a 256\,MiB cache and 256\,MiB of SQL memory. Vector indexes use cosine distance with the database's default partition sizing, and the vector-index feature is enabled cluster-wide. Rows are loaded as single-row transactions from four client threads through the gateway of the row's home region. The client is Python with psycopg~3 on the same host. Every strategy and gateway gets 15 warm-up queries before timing. The scripts, raw measurements, and the generator that emits every number in this paper from those measurements are available from the authors on request.

\textbf{Data and embeddings.} We use all \NumPassages{} context passages of SQuAD v1.1 (train and dev)~\cite{rajpurkar2016squad}. Each Wikipedia article is assigned a deterministic home region by hashing its title (\NumEast{} / \NumWest{} / \NumEU{} passages in us-east1 / us-west1 / europe-west1), modeling a tenant- or jurisdiction-partitioned corpus. Passages (title and text) and questions are embedded with WordLlama~\cite{wordllama}, a 256-dimensional static embedding model that runs on CPU. We L2-normalize the vectors, store them in a 256-dimensional vector column, and index them with the database's native partition-based cosine index (search beam 32 unless stated). S1--S3 use one table holding the full corpus. S4/S5 use one region-pinned table per region (Section~\ref{sec:discussion} explains why we did not use row-level partitioning for reads). With exact search over the full corpus, the answer-bearing passage appears in the top-10 for \ExactHitTen{} of questions. This is the quality ceiling for the embedding model; we report all quality numbers relative to it.

\textbf{Workload and metrics.} We draw a fixed random sample of \NumQSample{} SQuAD questions. We execute each of the first \NumBenchQ{} from \emph{all three} gateways under all six strategy variants, in shuffled order after warm-up (\NumBenchRuns{} timed top-10 queries). Queries run one at a time, so latency excludes queuing. We report client-observed latency (p50/p95), \emph{answer hit@10} (the gold passage is retrieved), and \emph{recall@10} against exact global top-10 neighbors. To model locality we define the skew $\sigma$: a question originates at its passage's home region with probability $\sigma$ and otherwise at one of the two other regions uniformly. Because every question was executed from every region, metrics under any $\sigma$ are computed exactly by weighting, without resampling. To check run-to-run stability, we restarted the cluster and repeated all six strategies on the first \RepN{} questions. Every remote median was reproduced within \RepRemRel{} and every local median within \RepLocAbs\,ms.

\section{Results}
\label{sec:results}

\subsection{Retrieval latency by client region}
\begin{figure}[t]
\centering
\includegraphics[width=\columnwidth]{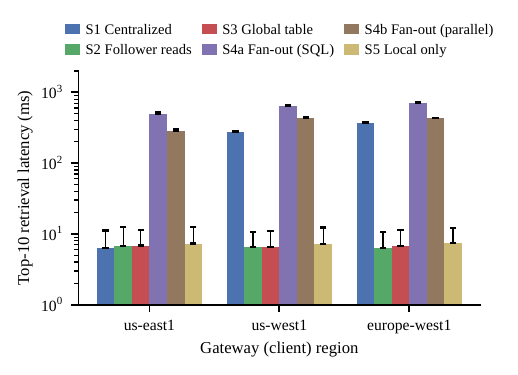}
\caption{Median top-10 latency (bars) and p95 (whiskers) by gateway region, log scale.}
\label{fig:lat}
\end{figure}

Table~\ref{tab:latency} and Fig.~\ref{fig:lat} report top-10 latency from each gateway. The strategies split into two groups with no middle ground. Every strategy that keeps the ANN search inside the client's region---S2, S3, S5, and S1 when the client is in us-east1---answers in \LocalMin--\LocalMax\,ms at the median. Every strategy that sends the search across the WAN costs hundreds of milliseconds. The spread between the fastest and slowest median is \MaxSpeedup.

\begin{table*}[t]
\centering
\caption{Top-10 retrieval latency (p50 / p95, ms) by client region, and retrieval quality averaged over all three client regions (600 questions per region).}
\label{tab:latency}
\footnotesize
\setlength{\tabcolsep}{6pt}
\begin{tabular}{@{}lccccc@{}}
\toprule
Strategy & us-east1 & us-west1 & europe-west1 & hit@10 & recall@10 \\
\midrule
S1 Centralized & 6.4 / 11.2 & 276.3 / 282.4 & 372.6 / 379.1 & 65.3\% & 90.4\% \\
S2 Follower reads & 6.7 / 12.5 & 6.5 / 10.7 & 6.4 / 10.6 & 65.3\% & 90.4\% \\
S3 Global & 6.9 / 11.3 & 6.6 / 10.9 & 6.7 / 11.4 & 66.3\% & 90.5\% \\
S4a Fan-out (SQL) & 500.9 / 522.3 & 644.4 / 660.6 & 716.6 / 732.5 & 70.5\% & 97.3\% \\
S4b Fan-out (parallel) & 285.4 / 298.1 & 429.0 / 440.1 & 429.0 / 438.7 & 70.5\% & 97.3\% \\
S5 Local only & 7.3 / 12.6 & 7.1 / 12.3 & 7.4 / 12.2 & 25.8\% & 32.4\% \\
\bottomrule
\end{tabular}
\end{table*}

\textbf{ANN amplifies WAN latency.} A primary-key read of the us-east1-homed table takes \CalEUPoint\,ms from europe-west1, about one RTT. The same client's top-10 vector query against that table (S1) takes \LatCenEU\,ms, so the remote search spends \CentralEUoverRTT{} RTTs beyond its local cost (\CentralWestoverRTT{} RTTs from us-west1). The reason is that the index search is a chain of dependent reads: root and interior partitions, leaf partitions, then a lookup join that fetches full-precision vectors for re-ranking. Each step waits for the previous one. Placement that is ``only one RTT away'' for OLTP is therefore several RTTs away for retrieval. Ingest shows the same effect: a single-row insert into the centralized table costs \IngCenWest\,ms from us-west1 and \IngCenEU\,ms from europe-west1, because index maintenance runs its own search.

\textbf{Fan-out multiplies it.} Searching all three home partitions in one SQL statement (S4a) costs \LatFanEast--\LatFanEU\,ms. This is close to the sum of the per-region searches, which suggests that the subqueries run one after another. Issuing the three searches concurrently from the client (S4b) reduces this to \LatParEast--\LatParEU\,ms, which is bounded by the slowest remote region. Even S4b is \ParOverLocMin--\ParOverLocMax$\times$ slower than the local strategies.

\textbf{Follower reads and global tables are local.} S2 and S3 serve the search from a local non-voting replica at \LatFrEU\ and \LatGloEU\,ms in Europe. Their costs are elsewhere. S2's results lag the latest write by a median of \FrLagSec\,s in our runs: a newly inserted passage became visible to a follower-read query \FrLagSec\,s after commit in every region. S3 makes every write wait out a commit-wait period, \IngGloEast--\IngGloEU\,ms per insert. Both keep a full copy of the corpus in every region. Home-partitioned writes are local (\IngPartEU--\IngPartEast\,ms into a per-region table). With row-level regional partitioning and a primary key that omits the region, each insert also runs a cross-region uniqueness check (\IngRbrEast--\IngRbrWest\,ms).

\subsection{Retrieval quality}
Strategies over the same single index (S1--S3) return nearly identical results: hit@10 is \HitCen--\HitGlo{} and recall@10 against exact search is \RecCen--\RecGlo. The partitioned fan-outs reach \HitPar{} hit@10 and \RecPar{} recall. This gain is \emph{not} a property of placement. Each of the three smaller indexes is searched with the default beam, so the fan-out explores more partitions in total. A single index reaches comparable quality with a larger beam. Raising the beam from 32 to 256 lifts recall@10 from \BeamDefRec{} to \BeamMaxRec{} and hit@10 from \BeamDefHit{} to \BeamMaxHit{}, while local median latency grows from \BeamDefLat{} to \BeamMaxLat\,ms. The more striking quality result is S5. For questions issued from the answer's home region, searching only the local partition yields \HitHomeLoc{} hit@10. That exceeds global search (\HitPar) and even exact search over the whole corpus (\ExactHitTenBench{} on the same questions), because the partition acts as a relevance filter that removes distractors homed elsewhere. For questions issued elsewhere, S5 cannot find the answer at all, so its average over all client regions is only \HitLoc.

\subsection{Locality skew and LARR}
\label{sec:skew}
\begin{figure}[t]
\centering
\includegraphics[width=\columnwidth]{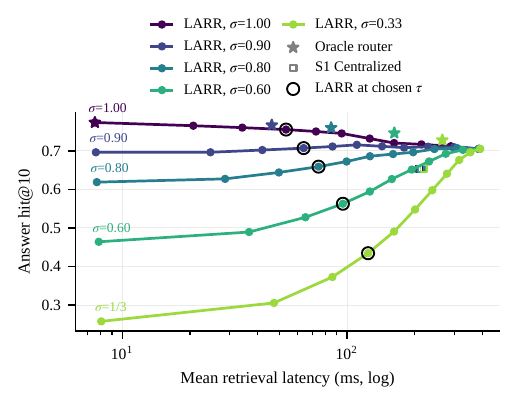}
\caption{Answer hit@10 vs mean latency as LARR's threshold $\tau$ sweeps from local-only (left end of each curve) to always-escalate (right end), for several locality skews $\sigma$. Stars: an oracle router that escalates exactly when the answer is homed elsewhere. Squares: centralized S1.}
\label{fig:larr}
\end{figure}

Table~\ref{tab:skew} and Fig.~\ref{fig:larr} combine latency and quality under locality skew $\sigma$. We use the threshold $\tau=\LarrTau$ (the \LarrTauPct{}th percentile of local top-1 distances). It is the lowest-latency setting whose hit@10 at $\sigma=0.9$ matches the global parallel fan-out. At $\sigma=0.9$, LARR escalates \LarrEscNine{} of questions. It matches the fan-out's quality (\LarrHitNine{} vs.\ \SgHitParNine) at a mean of \LarrLatNine\,ms instead of \SgLatParNine\,ms, a \LarrSpeedupNine{} reduction. A paired bootstrap over questions (2{,}000 resamples) puts the hit@10 difference at \CIHitDiff{} points (95\% CI), so the two are statistically indistinguishable in quality, while the speedup's 95\% CI is \CISpeedup$\times$. It also beats local-only search (\SgHitLocNine) while keeping every passage stored only in its home region. At $\sigma=1$, LARR reaches \LarrHitOne{}, between local-only and global search.

\begin{table*}[t]
\centering
\caption{Mean latency (ms) and answer hit@10 under locality skew $\sigma$ (fraction of questions issued from the region that homes the answer). LARR uses $\tau=0.611$.}
\label{tab:skew}
\footnotesize
\setlength{\tabcolsep}{7pt}
\begin{tabular}{@{}lccccc@{}}
\toprule
Strategy & $\sigma$=1.0 & 0.9 & 0.8 & 0.6 & 1/3 \\
\midrule
S1 Centralized & 210 / 65.3 & 211 / 65.3 & 213 / 65.3 & 216 / 65.3 & 219 / 65.3 \\
S2 Follower reads & 7 / 65.3 & 7 / 65.3 & 7 / 65.3 & 7 / 65.3 & 7 / 65.3 \\
S3 Global & 8 / 66.3 & 8 / 66.3 & 8 / 66.3 & 8 / 66.3 & 8 / 66.3 \\
S4b Fan-out & 378 / 70.5 & 379 / 70.5 & 380 / 70.5 & 382 / 70.5 & 384 / 70.5 \\
S5 Local only & 8 / 77.3 & 8 / 69.6 & 8 / 61.9 & 8 / 46.4 & 8 / 25.8 \\
\midrule
LARR (seq.) & 54 / 75.5 & 64 / 70.7 & 75 / 65.9 & 96 / 56.3 & 124 / 43.4 \\
LARR (hedged) & 52 / 75.5 & 63 / 70.7 & 73 / 65.9 & 94 / 56.3 & 121 / 43.4 \\
Oracle router & 8 / 77.3 & 46 / 76.7 & 85 / 76.0 & 162 / 74.6 & 265 / 72.8 \\
\bottomrule
\multicolumn{6}{@{}l}{\footnotesize Cells: mean latency (ms) / hit@10 (\%).}
\end{tabular}
\end{table*}

\textbf{Where LARR falls short.} As locality weakens, LARR's quality falls toward local-only search: \LarrHitUni{} at $\sigma=1/3$ with \LarrEscUni{} escalations. The cause is the confidence signal. Measured as a classifier of ``the answer is homed here,'' the local top-1 distance has an AUC of only \AucDone{} with our static embedding model. The oracle router in Table~\ref{tab:skew} shows the headroom: it reaches \OrHitNine{} at \OrLatNine\,ms for $\sigma=0.9$ and \OrHitUni{} at $\sigma=1/3$. A better escalation signal could recover much of this gap. Features of the local result alone do not provide one: a logistic model over five features of the local top-10 distances (top-1, top-1/top-2 margin, top-1/top-10 spread, mean, and standard deviation), fitted on questions 0--299, reached a held-out AUC of \AucLocalFeat{} on questions 300--599, versus \AucDOneTest{} for $d_1$ alone. The missing information is about the \emph{other} regions, which motivates LARR-S. The hedged variant saves only 1--3\,ms on average, because the local search is far cheaper than the remote one, so $\max(T_{\text{loc}},T_{\text{glob}})\approx T_{\text{loc}}+T_{\text{glob}}$.

\textbf{Live, held-out validation.} The threshold was chosen on the benchmark questions. To check that the result generalizes, we ran LARR end to end (local query, then a live parallel fan-out on escalation) with $\tau=\LarrTau$ on \LiveNQ{} \emph{held-out} questions from all three gateways. Weighted to $\sigma=0.9$, the live run escalated \LiveEscNine{} of questions and achieved \LiveHitNine{} hit@10 at a mean of \LiveLatNine\,ms. The offline estimate on the benchmark questions was \LarrEscNine{}, \LarrHitNine{}, and \LarrLatNine\,ms.

\begin{table}[t]
\centering
\caption{Held-out comparison of escalation signals: best hit@10 (\%) within a mean-latency budget, LARR ($d_1$ only) / LARR-S ($d_1$ + centroid sketch, $K{=}256$). Fitted on questions 0--299, evaluated on 300--599.}
\label{tab:sketch}
\footnotesize
\setlength{\tabcolsep}{4pt}
\begin{tabular}{@{}lccc@{}}
\toprule
$\sigma$ & $\le$50\,ms & $\le$100\,ms & $\le$200\,ms \\
\midrule
0.9 & 70.0 / 70.6 & 71.4 / 71.0 & 71.4 / 71.1 \\
0.8 & 63.6 / 65.8 & 66.3 / 68.4 & 69.3 / 70.3 \\
0.6 & 49.5 / 51.3 & 54.5 / 57.8 & 64.5 / 66.6 \\
1/3 & 28.9 / 31.2 & 37.7 / 40.6 & 52.9 / 56.2 \\
\bottomrule
\end{tabular}
\end{table}

\textbf{LARR-S on held-out questions.} We fit the LARR-S logistic model and both routers' thresholds on questions 0--299 and evaluate on questions 300--599 using the measured per-query latencies. The sketch raises the held-out AUC for ``the answer is homed here'' from \AucDOneTest{} ($d_1$ alone) to \AucSkB{} ($K{=}256$; \AucSkA{} with $K{=}64$, \AucSkC{} with $K{=}1024$); the sketch margin alone reaches \AucSketchOnly. Table~\ref{tab:sketch} shows what this buys end to end. At $\sigma \le 0.8$, LARR-S gains up to \SkGainMax{} points of hit@10 for the same latency budget. At $\sigma=0.9$, where $d_1$ alone already matches the fan-out, the two are within half a point. The remaining gap to the oracle comes from the fan-out itself, which returns the answer only \HitPar{} of the time. Routing an escalated query only to the region the sketch ranks highest, instead of to every region, is a natural next step that would also cut escalated latency.

\subsection{Production-style replication: three nodes per region}
\label{sec:ninenode}
The one-node-per-region testbed keeps one voting replica per range. To check that this simplification does not drive our conclusions, we rebuilt the corpus on a second cluster with three nodes in each region (nine nodes) and the database's default zone-level survivability: three voting replicas in the home region and non-voting replicas in the other two regions. We confirmed this placement for every table. We then reran S1, S2, S4a, S4b and S5 on the first 200 questions from all three gateways, along with single-row ingest. The conclusions hold:
\begin{itemize}
\item \emph{WAN amplification is unchanged.} A remote S1 search costs \NineCenEU\,ms from europe-west1 and \NineCenWest\,ms from us-west1. That is \NineAmpEU{} and \NineAmpWest{} RTTs beyond the local cost, versus \OneAmpEUsub{} on the same questions with one node per region. The parallel fan-out costs \NineParMin--\NineParMax\,ms.
\item \emph{Local strategies stay local, but slow down on our host.} S2 and S5 answer in \NineLocMin--\NineLocMax\,ms, up from roughly 7\,ms. Nine database processes share two vCPUs, so this increase reflects host CPU contention, not the network.
\item \emph{LARR's advantage is preserved.} At $\sigma=0.9$ and the same $\tau$, LARR reaches \NineLarrHit{} hit@10 versus \NineParHit{} for the parallel fan-out, at \NineLarrLat\,ms versus \NineParLat\,ms (\NineSpeedup). On the same 200 questions with one node per region, the figures were \OneLarrHitSub{} versus \OneParHitSub{} (\OneSpeedupSub).
\item \emph{Writes now pay an intra-region quorum.} A single-row insert into the writer's own region takes \NineIngLocMin--\NineIngLocMax\,ms, up from \IngPartEU--\IngPartEast\,ms. An insert into the centralized table from europe-west1 still takes \NineIngCenEU\,ms.
\end{itemize}

\subsection{Behavior under concurrent load}
\label{sec:concurrency}
\begin{figure}[t]
\centering
\includegraphics[width=\columnwidth]{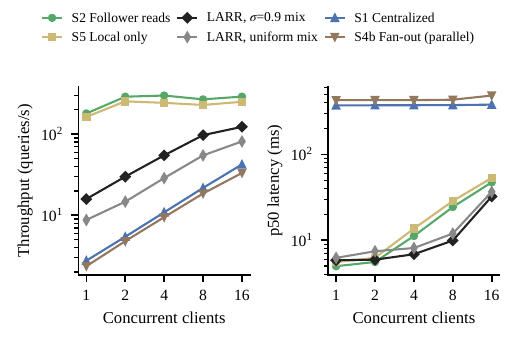}
\caption{Throughput (left) and median latency (right) versus closed-loop concurrency at the europe-west1 gateway, 25\,s per point, log scales.}
\label{fig:conc}
\end{figure}

Fig.~\ref{fig:conc} drives the europe-west1 gateway with 1 to 16 closed-loop clients for 25\,s per setting. The strategies fall into two regimes.
\begin{itemize}
\item \emph{Local strategies (S2, S5) are compute-bound.} They saturate at about \CcLocPeak{} queries/s on our two-vCPU host with only two to four clients. After that, latency grows with queueing.
\item \emph{Remote strategies (S1, S4b) are latency-bound.} Their throughput grows linearly with clients because each client waits on a chain of WAN round trips. From \CcSOneOne{} queries/s with one client, S1 reaches only \CcSOneSixteen{} with 16 clients. It would need about \CcSOneClientsHundred{} concurrent connections per 100 queries/s, a real cost in connection and memory capacity.
\item \emph{LARR sits between the two.} With the $\sigma=0.9$ question mix, it serves \CcLarrSixteen{} queries/s at 16 clients with a median of \CcLarrSixteenPfifty\,ms. With a uniform mix ($\sigma=1/3$) it serves \CcLarrUniSixteen{}.
\end{itemize}
LARR's p95 (\CcLarrSixteenPnf\,ms at 16 clients) is set by its escalated queries, which wait for the fan-out. LARR therefore improves the median and throughput, not the tail. Workloads with tight tail-latency targets should pair it with the hedged variant or with a per-query timeout that returns the local answer.

\subsection{A second embedding model}
\label{sec:minilm}
To test whether our conclusions depend on WordLlama, we re-embedded the corpus and questions with all-MiniLM-L6-v2~\cite{reimers2019sbert}, a 384-dimensional transformer sentence encoder. We then loaded the vectors into a second set of S1 and per-region tables with their own vector indexes. With this model, exact search places the answer passage in the top-10 for \MlExact{} of questions, against \ExactHitTen{} for WordLlama. We ran S1, S2, S4b and S5 on \MlNQ{} questions from all three gateways. We re-calibrated LARR's threshold on the first half of the questions and evaluated it on the second half. On the calibration half, the paper's rule produced a near-tie: the best threshold reached \MlCalBest{} against \MlCalFan{} for the fan-out. We therefore applied it with a one-point non-inferiority margin, which selects $\tau$ at the \MlTauPct{}th percentile of local top-1 distances.
\begin{itemize}
\item \emph{Placement effects are model-independent.} The remote S1 search from europe-west1 takes \MlCenEU\,ms, \MlAmp{} RTTs beyond its local cost. Local strategies answer in \MlLocMin--\MlLocMax\,ms.
\item \emph{Quality rises with the better model, but the ranking of strategies does not change.} Hit@10 is \MlHitCen{} for S1, \MlHitPar{} for the parallel fan-out, and \MlHitHomeLoc{} for local-only search on home-region questions.
\item \emph{LARR's result carries over, with a smaller speedup.} On the held-out half at $\sigma=0.9$, LARR reaches \MlLarrHit{} hit@10 against \MlParHit{} for the fan-out, at \MlLarrLat\,ms instead of \MlParLat\,ms (\MlSpeedup), escalating \MlLarrEsc{} of questions. A threshold at the median distance trades a fraction of a point (\MlMedHit) for a \MlMedSpeedup{} speedup (\MlMedLat\,ms). The speedup is smaller than with WordLlama because the stronger model gets more out of the remote partitions, so matching the fan-out takes more escalations.
\item \emph{The confidence signal is stronger.} The local top-1 distance has an AUC of \MlAucDone{}, against \AucDOneTest{} for WordLlama. With the centroid sketch added it reaches \MlAucSketch{}.
\end{itemize}

\subsection{Corpus size}
\label{sec:scale}
To see how the round-trip multiplier grows with the index, we split every passage into sentences (\ScChunks{} title-prefixed chunks) and loaded nested random subsets of 10k, 30k, and 100k chunks into three us-east1-pinned tables, each with its own vector index. We ran 200 questions against each from all three gateways. A remote top-10 search costs \ScMultA{} RTTs beyond its local cost at 10k vectors (\ScEUA\,ms from europe-west1), \ScMultB{} at 30k (\ScEUB\,ms), and \ScMultC{} at 100k (\ScEUC\,ms). Over the same range, local latency grows only from \ScLocA{} to \ScLocC\,ms. Measured against calibrated point reads, which include the proxy's few milliseconds per round trip, these are \ScMeasA{}, \ScMeasB{} and \ScMeasC{} round trips. The multiplier moves in whole round trips. That pattern is consistent with one extra dependent read for each extra level of the partition tree, so the multiplier should grow roughly with the logarithm of the corpus size, not linearly. We expect multi-million-vector corpora to add one or two more levels and so one or two more RTTs per remote search, which strengthens the case for local-first retrieval.

\subsection{Real cloud regions and WAN sensitivity}
\label{sec:wan}
\begin{table}[t]
\centering
\caption{Real cloud regions (600 questions; RTTs measured to us-east1) and emulated WAN profiles (200 questions). Ampl.: remote S1 cost beyond local, in measured RTTs. LARR/fan: mean ms at $\sigma=0.9$.}
\label{tab:wan}
\footnotesize
\setlength{\tabcolsep}{3pt}
\begin{tabular}{@{}lccccc@{}}
\toprule
Profile & RTTs (ms) & S1 EU p50 & Ampl. & LARR/fan & Gain \\
\midrule
Real cloud & 66/100/-- & 424 & 4.2--4.3 & 75 / 458 & 6.1$\times$ \\
\midrule
Emulated & 64/88/136 & 373 & 4.0 & 64 / 378 & 5.9$\times$ \\
+ jitter & 64/88/136 & 415 & 4.1--4.2 & 72 / 432 & 6.0$\times$ \\
Nearby & 12/18/26 & 95 & 4.1--4.2 & 23 / 95 & 4.2$\times$ \\
\bottomrule
\end{tabular}
\end{table}

\textbf{Real cloud.} We repeated the core measurements on three public-cloud VMs (\CwType, 2 vCPUs) in the real regions us-east1, us-west1 and europe-west1, connected over the provider's inter-region network, with the same placement and \CwNQ{} questions from every region. Point reads measured RTTs of \CwRttWest\,ms and \CwRttEU\,ms to us-east1. A remote S1 search costs \CwAmpWest{} and \CwAmpEU{} real RTTs beyond its local cost (\CwCenEU\,ms from europe-west1); local strategies take \CwLocMin--\CwLocMax\,ms. At $\sigma=0.9$ and the paper's $\tau$, LARR reaches \CwLarrHit{} hit@10 against \CwFanHit{} for the parallel fan-out, at \CwLarrLat\,ms instead of \CwFanLat\,ms (\CwSpeedup). Live on \CwLiveN{} held-out questions, LARR reaches \CwLiveHit{} at \CwLiveLat\,ms. The real network shows up in the tails: the fan-out from us-west1 has a p95 of \CwFanPnfWest\,ms against a median of \CwFanPfWest\,ms.

\textbf{Emulator sensitivity.} Changing only the delay proxies, we re-ran S1, S5 and S4b on \WnNQ{} questions with jitter (an extra exponentially distributed delay per chunk, mean \WnJitPct\% of the one-way delay) and with nearby-region RTTs of 12/18/26\,ms. Amplification stays at \WnAmpMin--\WnAmpMax{} RTTs, LARR's speedup is \WnSpeedJit{} and \WnSpeedNear, and result sets are \WnHitSame{} across profiles.

\section{Discussion}
\label{sec:discussion}
\textbf{Design guidance.} Our measurements suggest four rules for placing a RAG corpus in a multi-region SQL database.
(1)~\emph{Never serve interactive retrieval from a remote leaseholder.} The ANN traversal amplifies the WAN RTT several times over, unlike a point read.
(2)~If the corpus may be copied to every region and a few seconds of staleness is acceptable, \emph{follower reads} on a single-homed table give local latency for free. A global table gives the same read latency without staleness, at the cost of slow writes.
(3)~If passages must stay in their home region, \emph{partition by home region} and search locally first. A global fan-out should be the exception, not the default, and it should be issued in parallel rather than as one SQL statement.
(4)~Calibrate LARR's threshold on the local top-1 distance distribution of the deployed embedding model. A single scalar knob then trades latency for completeness.

\textbf{End-to-end RAG impact.} Retrieval sits on the critical path before generation starts, so every millisecond LARR saves comes straight off time-to-first-token. At $\sigma=0.9$, that is about \LarrSaveMs\,ms per question against the parallel fan-out, and the generation time itself is unchanged. Answer quality is bounded by whether the answer-bearing passage reaches the prompt. LARR matches the fan-out's hit@10 within the confidence interval, so the generator sees the gold passage equally often. We did not measure generated-answer quality, which also depends on the LLM and the prompt. We leave that end-to-end evaluation to future work.

\textbf{Larger and changing corpora.} Larger corpora raise the remote penalty (Section~\ref{sec:scale}), which favors local-first designs. Frequent updates also favor home-region partitioning: in our measurements a home-region insert stays local (\IngPartEU--\IngPartEast\,ms), while a centralized or global table charges every remote writer WAN round trips or a commit wait. Per-document access rules are filtered ANN queries~\cite{gollapudi2023filtered,patel2024acorn}, and LARR composes with them: the same filter applies to the local search and to any escalation.

\textbf{A pitfall with partitioned vector indexes.} Our first S4/S5 implementation used a row-level regionally partitioned table with a vector index. Relying on the implicit region partitioning, top-$k$ searches returned incorrect neighbors on two recent releases; even a passage's own embedding was not its nearest neighbor. An explicit region prefix fixed a small sequential test but failed again after a concurrent bulk load, while identically loaded non-partitioned indexes passed. We therefore used one region-pinned table per region, which gives the same placement. Practitioners should run a self-retrieval check (each stored vector should return itself at rank 1) after bulk-loading any partitioned vector index.

\textbf{Threats to validity.} The WAN is emulated in user space on a single 2-vCPU host, so absolute local latencies are pessimistic, and the proxy adds a few milliseconds to each cross-region round trip (Section~\ref{sec:setup}). A real three-region cloud deployment reproduces the amplification and LARR's speedup (Section~\ref{sec:wan}); it used one cloud provider and one node per region. The main testbed uses one node per region with one voter. A three-nodes-per-region rerun (Section~\ref{sec:ninenode}) reproduces the read-path conclusions and quantifies the added intra-region write quorum. Our main embedding model, WordLlama, is a small static model. A transformer encoder raises absolute quality but leaves the placement conclusions unchanged (Section~\ref{sec:minilm}). LARR's benefit depends on how well the local top-1 distance separates answerable from unanswerable local searches, and that separation is model-dependent. The corpus has \NumPassages{} passages. Larger partitions deepen the index tree, which increases the number of dependent reads per query and would \emph{amplify} the WAN effects we report. Region assignment is synthetic (hashing article titles), and $\sigma$ is a model of locality rather than a measured trace. An enterprise trace with natural geographic locality would let $\sigma$ be measured rather than swept.

\section{Conclusion}
Vector search inside a geo-distributed SQL database inherits that database's placement options, but it does not inherit the latency intuitions built on point reads. In a three-region geo-distributed SQL testbed, a remote ANN query cost \CentralEUoverRTT{} WAN round trips. Local strategies answered in \LocalMin--\LocalMax\,ms, while remote and fan-out strategies took hundreds of milliseconds. When data must stay in its home region, a simple distance-threshold router (LARR) recovers most of the quality of a global search at a fraction of its latency under realistic locality. Future work includes routing escalations only to the region the sketch ranks highest, LARR for hybrid (vector + relational) predicates, larger real-cloud deployments, and support for access-residency policies during escalation.

\bibliographystyle{IEEEtran}
\bibliography{refs}

\end{document}